\PassOptionsToPackage{table}{xcolor}
\documentclass[]{ceurart}
\usepackage{cmap} 
\usepackage[T1]{fontenc}
\usepackage{underscore}           

\usepackage{natbib}
\usepackage{amssymb}

\usepackage{amsmath}
\usepackage{amssymb}
\usepackage{stmaryrd}
\usepackage{commath}
\usepackage{scalerel}
\usepackage{pdflscape}
\usepackage[table]{xcolor}
\DeclareMathOperator{\id}{\mathit{id}}
\DeclareMathOperator{\aand}{\wedge}
\newcommand{\D}[0]{\ensuremath{\mathbb{D}}}
\newcommand{\R}[0]{\ensuremath{\mathbb{R}}}
\newcommand{\m}[0]{\ensuremath{\mathbb{M}}}
\newcommand{\compl}[1]{\overline{#1}}
\newcommand{\inr}[1]{\text{in}_r({#1})}
\newcommand{\inl}[1]{\text{in}_l({#1})}
\newcommand{\tot}[1]{\mathbb{T}({#1})}
\newcommand{\Proj}[2]{\text{Proj}_{#1}({#2})}
\newcommand{\powerset}[1]{\mathbf{2}^{#1}}
\DeclareMathOperator{\union}{\cup}
\DeclareMathOperator{\taggedUnion}{\dot{\cup}}
\usepackage{newunicodechar}
\newunicodechar{⦷}{\text{\textcircled{\scaleto{\ensuremath{\parallel}}{0.8em}}}}
\DeclareMathOperator{\parProduct}{⦷}
\DeclareMathOperator{\disjProduct}{\otimes}
\DeclareMathOperator{\reconsProduct}{\oplus}
\DeclareMathOperator{\join}{\bowtie}

\title{Data accounting and error counting}

\author[1]{Micha\l{} J. Gajda}[%
orcid=0000-0001-7820-3906,
email={mjgajda@migamake.com},
url={https://migamake.com},
]
\address[1]{Migamake Pte Ltd}

\copyrightyear{2022}
\copyrightclause{Copyright for this paper by its authors.}

\conference{Arxiv}

\usepackage{xurl}
\usepackage{longtable,booktabs,tabularx}
\let\toprule\relax
\let\bottomrule\relax
\let\midrule\relax

\usepackage{tikz-cd}

\IfFileExists{footnote.sty}{\usepackage{footnote}\makesavenoteenv{longtable}}{}
\usepackage{listings}

\providecommand{\tightlist}{%
  \setlength{\itemsep}{0pt}\setlength{\parskip}{0pt}}

\newenvironment{Shaded}{}{}
\usepackage{fancyvrb}

\newcommand{\KeywordTok}[1]{\textcolor[rgb]{0.00,0.00,1.00}{#1}}
\newcommand{\NormalTok}[1]{#1}
\newcommand{\OperatorTok}[1]{#1}
\newcommand{\OtherTok}[1]{\textcolor[rgb]{1.00,0.25,0.00}{#1}}

\DefineVerbatimEnvironment{Highlighting}{Verbatim}{commandchars=\\\{\}}

\newlength{\cslhangindent}
\newlength{\csllabelwidth}
\newenvironment{CSLReferences}[2] 
 {
  \setlength{\parindent}{0pt}
  \ifodd #1 \everypar{\setlength{\hangindent}{\cslhangindent}}\ignorespaces\fi
  \ifnum #2 > 0
  \setlength{\parskip}{#2\baselineskip}
  \fi
 }%
 {}
\usepackage{calc} 

\begin{document}
\maketitle

\begin{abstract}
Can we infer sources of errors from outputs of the complex data
analytics software? Bidirectional programming promises that we can
reverse flow of software, and translate corrections of output into
corrections of either input or data analysis.

This allows us to achieve holy grail of automated approaches to
debugging, risk reporting and large scale distributed error tracking.

Since processing of risk reports and data analysis pipelines can be
frequently expressed using a sequence relational algebra operations, we
propose a replacement of this traditional approach with a data
summarization algebra that helps to determine an impact of errors. It
works by defining data analysis of a necessarily complete summarization
of a dataset, possibly in multiple ways along multiple dimensions. We
also present a description to better communicate how the complete
summarizations of the input data may facilitates easier debugging and
more efficient development of analysis pipelines. This approach can also
be described as an generalization of axiomatic theories of accounting
into data analytics, thus dubbed data accounting.

We also propose formal properties that allow for transparent assertions
about impact of individual records on the aggregated data and ease
debugging by allowing to find minimal changes that change behaviour of
data analysis on per-record basis.
\end{abstract}

\begin{keywords}
reversible computation
\sep
data analytics
\sep
error estimation
\sep
accounting measure
\sep
time-invariant measure
\sep
relational algebra
\end{keywords}

\hypertarget{introduction}{%
\section{Introduction}\label{introduction}}

Error tracing and error-impact estimation has been a holy grail of
practical use of automated approaches to risk reporting(Basel Committee
on Banking Supervision 2013), large-scale distributed error
tracking(Julian 2017), automated debugging.

Since processing of risk reports and data analysis pipelines can be
frequently expressed using a sequence relational algebra operations, we
propose a replacement of this traditional approach with a data
summarization algebra that helps to determine an impact of errors. It
works by defining data analysis of a necessarily complete summarization
of a dataset, possibly in multiple ways along multiple dimensions.

We also present a description to better communicate how the complete
summarizations of the input data may facilitates easier debugging and
more efficient development of analysis pipelines.

This approach can also be described as an generalization of axiomatic
theories of accounting into data analytics, thus dubbed data accounting.

We also propose formal properties that allow for transparent assertions
about impact of individual records on the aggregated data and ease
debugging by allowing to find minimal changes that change behaviour of
data analysis on per-record basis.

The answer is indeed based on fully bidirectional computation and
generalization of it to data-preserving transformations.

\hypertarget{literature-review}{%
\section{Literature review}\label{literature-review}}

Basel Committee Best Practices (BCBS) 239(Basel Committee on Banking
Supervision 2013) is a set of recommendations for the financial
institutions on how to organize their risk reporting pipelines. These
recommendations also give a careful considerations to general principles
of data aggregation and reporting, and thus became reference for other
work on best practice in aggregated reporting.

We also are inspired by practical rules coming from recent opinions on
data pipeline construction methodology(Gajda 2020a, 2020b).

\hypertarget{bidirectional-computing}{%
\subsection{Bidirectional computing}\label{bidirectional-computing}}

Reversible computing(Bennett 1973) aims to reverse computation from
outputs to inputs. Bidirectional computing(Bancilhon and Spyratos 1985)
attempts to generalize programs to allow picking a subset of either
outputs and inputs, and computing the remaining.

Automated debugging aims to automatically indicate possible places in
the program that may be buggy. Automated program derivation aims to
propose valid programs given both inputs and outputs.

Differentiable computing(Levitt 1983; Abadi and Plotkin 2019) aims to
find a minimum of one function, by computing both its value and
differential on the pivot point.

\hypertarget{database-view-update-problem}{%
\subsection{Database view update
problem}\label{database-view-update-problem}}

Database view is a computation from input relations to output relation.
Database community have long tried to resolve \emph{view update
problem}(Chen and Liao 2011) which tries to reverse computation of the
view, so that the update on results is transformed into update on source
relations.

While this approach was limited in scope to relations, it has great
impact on data analytics, since big datasets are often modeled directly
as relations, or translatable to relations (for example graphs
translated into incidence relation).

Analysis of the problem has led to identification of \emph{view
complement} which contains relation missing from the view, but essential
for computing the update.

In the sense the database community identified \emph{views} as
non-injective. In order to achieve updatable views, it attempted to
restore the injectiveness of the \emph{views} by computing \emph{view
complement}.

\hypertarget{event-sourcing}{%
\subsection{Event sourcing}\label{event-sourcing}}

Event sourcing replaces database as a source of truth with a history of
events that happened. This \emph{event log} corresponds to general
ledger in accounting, in the sense that it can only be appended to, and
the state becomes just a \emph{view} or \emph{balance} on the ledger.

\hypertarget{automated-debugging}{%
\subsection{Automated debugging}\label{automated-debugging}}

Automated debugging approaches aim to find a location of a bug in the
program from the fault in the output. This may be realized with approach
akin to bidirectional program, but applied to program source, instead of
inputs.

\hypertarget{philosophy-and-axiomatization-of-accounting}{%
\subsection{Philosophy and axiomatization of
accounting}\label{philosophy-and-axiomatization-of-accounting}}

Accounting is a science that attempts to predict future performance
(Ijiri 2018, 1975) of companies by insight into cash flows and metrics.
The mathematical axiomatization of accounting(Ijiri 2018) is based on a
concept of \emph{time-invariant measure} that partitions money in a way
that changes along time dimension, but preserves a total sum.

This is consistent with a traditional observation of accountants that it
is better to have a \emph{monotonic} measure of cash flow instead of
recording only the sum of money currently available, and instead of
adding and removing money, move the money between different
\emph{accounts}.

This approach, called double entry accounting(Ellerman 2007), is best
presented by presenting a total balance as pair of \emph{debit} and
\emph{credit} sides. Subtraction of \emph{debit} from \emph{credit}
yields a total balance. Both sides are non-negative, and strictly
monotonic over time. That means that we have a strictly monotonic
measure that is easy to verify across time.

Next step is generalization of double-entry accounting to multiple
\emph{purpose accounts}: for example, we can make separately make
account for cash balance, spending on

In order to deal with difficulty of putting all elements of the
bussiness on monetary scale, modern accountants base statements not just
on ledger of past transactions, but \emph{infobase} defined as entire
database used to define financial statements(Warsono, Ridha, and
Darmawan 2009).

Note that valuation-centred approach (like \emph{fair value accounting}
-- FVA) makes it hard to communicate non-monetary information, and is
subject to currency stability. It also faces difficulty in face of
barter transactions where one company gives services directly in
exchange for another service, with no money changing hands (Balzer and
Mattessich 1991).

As opposed to FVA, traditional accounting is based on ``going-on
concern'' which values assets on non-liquidation basis, and thus is
oriented towards future performance of the company.

Information content school emphasises that disclosure of business
information attempts to reduce uncertainty regarding the entity. It sees
accounting information as a pro-active management support tool, that is
not centred on \textbf{fair value}.

\hypertarget{accounting-measure}{%
\subsubsection{Accounting measure}\label{accounting-measure}}

In the work on axioms of accounting measure
\href{https://www.degruyter.com/view/journals/ael/8/1/article-20170057.xml}{Ijiri}(Ijiri
2018) notes that all pertinent objects are quantifiable by amount,
volume, weight or another measure, and divided in a countable
collections of classes. For each class there must be a physical measure,
but they do not be the same for different classes. The principle of
substitubility tells that two sets of objects that have the same
quantity, can be mutually substituted for the purpose of exchanges.
Non-substitutible objects must be accounted in a class of itself.
Measures must be correct mathematical measures that are non-negative,
zero if class is empty, and countably additive (monotonic). Some objects
may possess multiple different classes that often correspond to
different dimensions of reporting. Time is a real variable over which
some objects change their classes. But total measure of objects over
time must be \emph{invariant}. Measures may be allocated. imputated and
compared, based on current and future value of an asset. Default is to
use cost of replacement by a given point of time.

For example, we can estimate a cost of purchase for a workstation at
1000\$ in 2020. The value of a workstation will decrease due to wear and
obsolescence to 800\$ 2021. But since the total measure must remain
1000\$, we divide this value in 2021 into two classes: current value of
a workstation at \$800, and written off value of \$200.

Balzer and Mattesich(Balzer and Mattessich 1991), and Ellerman(Ellerman
2007) made different attempts at axiomatizing value-based accounting
that partitions resources intro disjoint groups, but they go along
similar principles.

For example Ellerman(Ellerman 2007) gives a solution to monotonic
tracking of information contained in both cash inflows and outflows by
splitting them into pair of monotonically increasing values of
\emph{debit} and \emph{credit} used in traditional double-entry
accounting systems.

Some accounting philosophers add that the goal of accounting is more
than just to sum up the numbers, it is to predict future state of
business, and reduce uncertainty about this state(Warsono, Ridha, and
Darmawan 2009). This means that accounting becomes a science of data
analysis and summarization where accounting measure plays just a small
part within a big edifice of information summarization.

\hypertarget{requirements-for-error-impact-reporting}{%
\subsection{Requirements for error impact
reporting}\label{requirements-for-error-impact-reporting}}

Here we summarize basic requirements for error impact reporting in bulk:

The analytics pipeline does not \textbf{crash} for failing record, but
instead runs as a whole for a given algebra expression, and provides
\emph{correct relation(s)} (that contain correct and relevant results)
\textbf{and} \emph{error-trace relations} that complement correct
results with information needed for error reporting. That means that
functions \emph{must be total}.

\emph{No information is simply lost or discarded.} Whenever a relevant
information should not affect the output, we sort it to a
\textbf{different category} of information that is provided, but
irrelevant so it can be put into an error trace relation. We may convert
between types of information, but non-null information must be converted
to entities of non-null information. We treat information as monotonic
(it can only increase), and additive.

With every \emph{correct relation} that contains data that is valid and
relevant at the current stage of the pipeline we also keep an
\emph{error trace} with records that were discarded and thus they could
not impact the processing. This allows us to estimate the possible
impact of erroneous records, if they were provided on the input. This
means that we preserve erroneous inputs in the form of error outputs.

\hypertarget{railroad-oriented-programming}{%
\subsection{Railroad-oriented
programming}\label{railroad-oriented-programming}}

At the end of the analytics pipeline we should always be able to provide
an automatic \emph{error dashboard} that summarizes how much data was
correctly processed, and summarized in the output, and how much data was
discarded, and for what reasons. This again means that the
\emph{dashboard} that gives a final summary of the output must be
\emph{sensitive} to all inputs and all forms of errors.

We should automatically tag the error trace with all information
necessary to discover at which stage of the pipeline the data was
discarded. This means that we should divide the analytics pipeline into
summarization stages, and transformation stages, and transformation
stages should preserve sufficient information about bits of input to
evaluate their impact all the way until final summarization.

We conclude that the best approach to fulfill these requirements is to
attempt to keep all \textbf{errored records} along with the correct
records, and summarize them in parallel.

\textbf{Railroad-oriented programming} methodology(Wlaschin 2012)
suggests merging of all \textbf{error records} together in a separate
output stream that is also summarized, just like the data on the so
called \textbf{happy path} (main path that yields validated records for
summarization). This leads to two parallel streams of processing, one
forwarding valid data, the other forwarding errors and anomalies.

\begin{landscape}
\rowcolors{1}{gray!25}{white}

\begin{longtable}[]{@{}
  >{\raggedright\arraybackslash}p{(\columnwidth - 10\tabcolsep) * \real{0.1871}}
  >{\raggedright\arraybackslash}p{(\columnwidth - 10\tabcolsep) * \real{0.1742}}
  >{\raggedright\arraybackslash}p{(\columnwidth - 10\tabcolsep) * \real{0.1355}}
  >{\raggedright\arraybackslash}p{(\columnwidth - 10\tabcolsep) * \real{0.1032}}
  >{\raggedright\arraybackslash}p{(\columnwidth - 10\tabcolsep) * \real{0.2000}}
  >{\raggedright\arraybackslash}p{(\columnwidth - 10\tabcolsep) * \real{0.2000}}@{}}
\caption{Translation of shared concepts within literature of different
domains.}\tabularnewline
\toprule
\begin{minipage}[b]{\linewidth}\raggedright
Accounting
\end{minipage} & \begin{minipage}[b]{\linewidth}\raggedright
Bidirectional computing
\end{minipage} & \begin{minipage}[b]{\linewidth}\raggedright
Databases
\end{minipage} & \begin{minipage}[b]{\linewidth}\raggedright
Event sourcing
\end{minipage} & \begin{minipage}[b]{\linewidth}\raggedright
Data structures
\end{minipage} & \begin{minipage}[b]{\linewidth}\raggedright
Railroad-oriented programming
\end{minipage} \\
\midrule
\endfirsthead
\toprule
\begin{minipage}[b]{\linewidth}\raggedright
Accounting
\end{minipage} & \begin{minipage}[b]{\linewidth}\raggedright
Bidirectional computing
\end{minipage} & \begin{minipage}[b]{\linewidth}\raggedright
Databases
\end{minipage} & \begin{minipage}[b]{\linewidth}\raggedright
Event sourcing
\end{minipage} & \begin{minipage}[b]{\linewidth}\raggedright
Data structures
\end{minipage} & \begin{minipage}[b]{\linewidth}\raggedright
Railroad-oriented programming
\end{minipage} \\
\midrule
\endhead
double-entry accounting & reversible transformation & mark for deletion,
data partition & N/A & partition & switching between \emph{happy path}
and \emph{error path} \\
general ledger & & monotonic query & event log & append-only or
log-structured & N/A \\
N/A & bidirectional computation & view update problem & N/A & N/A &
N/A \\
sensitive to inputs & bijective transformation & injectivity & N/A & not
invariant, injectivity & N/A \\
\bottomrule
\end{longtable}

\end{landscape}

\hypertarget{proposed-work}{%
\section{Proposed work}\label{proposed-work}}

This work attempts to unify approaches of bidirectional programming,
railroad-oriented programming, and accounting to get the best features
of all three. That is we aim to provide both injectiveness of debugging
data pipelines (understood as finding difference to the program that
fixes it), and in order to do so, we maintain strict injectiveness of
view/filter computations, and monotonic data aggregations. In this
process, we discover that unifying theory for bidirectional computation
\textbf{and} debugging can be found on the basis of mathematical theory
of accounting.

To our knowledge bidirectional computing was so far not applied to
monotonic data aggregations. We also find that this is first work
exploring connection between axiomatic accounting(Ijiri 2018) and both
bidirectional computing and monotonic summarizations(Ross and Sagiv
1997).

\hypertarget{motivating-example}{%
\subsection{Motivating example}\label{motivating-example}}

We propose the following running example is that of the container ship
contents.

Container ship example table ``items'':

\begin{longtable}[]{@{}
  >{\raggedright\arraybackslash}p{(\columnwidth - 10\tabcolsep) * \real{0.1566}}
  >{\raggedleft\arraybackslash}p{(\columnwidth - 10\tabcolsep) * \real{0.1928}}
  >{\raggedright\arraybackslash}p{(\columnwidth - 10\tabcolsep) * \real{0.1566}}
  >{\raggedleft\arraybackslash}p{(\columnwidth - 10\tabcolsep) * \real{0.1928}}
  >{\raggedright\arraybackslash}p{(\columnwidth - 10\tabcolsep) * \real{0.0964}}
  >{\raggedright\arraybackslash}p{(\columnwidth - 10\tabcolsep) * \real{0.2048}}@{}}
\toprule
\begin{minipage}[b]{\linewidth}\raggedright
Description
\end{minipage} & \begin{minipage}[b]{\linewidth}\raggedleft
Purchase price
\end{minipage} & \begin{minipage}[b]{\linewidth}\raggedright
Commodity
\end{minipage} & \begin{minipage}[b]{\linewidth}\raggedleft
Quantity
\end{minipage} & \begin{minipage}[b]{\linewidth}\raggedright
Unit
\end{minipage} & \begin{minipage}[b]{\linewidth}\raggedright
Insurance
\end{minipage} \\
\midrule
\endhead
Sailors & \emph{priceless} & & 17 & person & 2,000,000\$ \\
Nutella & 10\$ & & 20,000,000 & jar & \\
Grain & & wheat grain & 200 & tonne & \\
Milk & & milk & 5,000 & litre & \\
Cat & & & 1 & animal & \\
Nut oil & \emph{unknown} & nut oil & 1,000 & litre & \\
Ship & 2,000,000,000\$ & & 1 & & 1,000,000,000\$ \\
\bottomrule
\end{longtable}

Commodity prices reference table ``Current prices'':

\begin{longtable}[]{@{}
  >{\raggedright\arraybackslash}p{(\columnwidth - 12\tabcolsep) * \real{0.1531}}
  >{\raggedleft\arraybackslash}p{(\columnwidth - 12\tabcolsep) * \real{0.1327}}
  >{\raggedleft\arraybackslash}p{(\columnwidth - 12\tabcolsep) * \real{0.1224}}
  >{\raggedleft\arraybackslash}p{(\columnwidth - 12\tabcolsep) * \real{0.1224}}
  >{\raggedleft\arraybackslash}p{(\columnwidth - 12\tabcolsep) * \real{0.2143}}
  >{\raggedright\arraybackslash}p{(\columnwidth - 12\tabcolsep) * \real{0.1224}}
  >{\raggedleft\arraybackslash}p{(\columnwidth - 12\tabcolsep) * \real{0.1327}}@{}}
\toprule
\begin{minipage}[b]{\linewidth}\raggedright
Market
\end{minipage} & \begin{minipage}[b]{\linewidth}\raggedleft
Commodity
\end{minipage} & \begin{minipage}[b]{\linewidth}\raggedleft
Price
\end{minipage} & \begin{minipage}[b]{\linewidth}\raggedleft
Currency
\end{minipage} & \begin{minipage}[b]{\linewidth}\raggedleft
Per unit
\end{minipage} & \begin{minipage}[b]{\linewidth}\raggedright
Quote date
\end{minipage} & \begin{minipage}[b]{\linewidth}\raggedleft
Future date
\end{minipage} \\
\midrule
\endhead
forward & wheat grain & 6.0981 & USD & bushel & 2021-12-01 &
2022-02-01 \\
spot & wheat grain & 6.0575 & USD & bushel & 2021-09-01 & \\
spot & wheat grain & \emph{closed} & USD & bushel & 2021-12-01 & \\
spot & milk & 33.98 & EUR & 100kg & 2021-12-01 & \\
\bottomrule
\end{longtable}

Table of reports:

\begin{longtable}[]{@{}
  >{\raggedleft\arraybackslash}p{(\columnwidth - 6\tabcolsep) * \real{0.1875}}
  >{\raggedright\arraybackslash}p{(\columnwidth - 6\tabcolsep) * \real{0.3672}}
  >{\raggedright\arraybackslash}p{(\columnwidth - 6\tabcolsep) * \real{0.3359}}
  >{\raggedright\arraybackslash}p{(\columnwidth - 6\tabcolsep) * \real{0.1094}}@{}}
\toprule
\begin{minipage}[b]{\linewidth}\raggedleft
Name
\end{minipage} & \begin{minipage}[b]{\linewidth}\raggedright
Description
\end{minipage} & \begin{minipage}[b]{\linewidth}\raggedright
Accounted
\end{minipage} & \begin{minipage}[b]{\linewidth}\raggedright
Unaccounted
\end{minipage} \\
\midrule
\endhead
Insured value & Total amount paid by insurer to all parties & Insured
value & Cat, nut oil \\
Replacement cost & Cost of replacement for owner in case of loss &
Purchase price or hiring cost & Cat, nut oil \\
Weight & Total mass transported & Weight of commodities & Sailors,
cat \\
Weight & Total mass transported & Weight of commodities & Sailors,
cat \\
\bottomrule
\end{longtable}

The point is to show how different value may be computed depending
whether we consider a total payment by insurer for the ship loss, cost
of replacing the ship by a ship owner, or a cost of purchasing
commodities on the market upon unreasonable delay. Since we will be
interested only in \emph{complete summarizations} of the dataset, we
will only build queries that summarize relevant data \emph{and summarize
omitted or ignored data}. This is considered the best practice(Basel
Committee on Banking Supervision 2013).

\hypertarget{outline}{%
\subsection{Outline}\label{outline}}

The similarity between best practices in impact reporting(Basel
Committee on Banking Supervision 2013) and mathematical theory of
accounting measure(Ijiri 2018) is not accidental. We may formalize
\emph{complete summarizations} of an input data set that generalize both
notion of error impact report, and accounting measure defined above.

But instead of \emph{measure} which is rather narrow numeric value, we
generalize to use a notion of information that is an additive monoid
over a given set. Notion of \emph{ownership} is translated into
\emph{partitions} of the set. Invariance of total measure over time is
replaced by invariance of total information monoid during processing.

Data space represents a complete summarization of a dataset. The
language that can only formulate complete summarizations would work
different from traditional relational algebra, and thus to define it we
need a mathematical theory.

\hypertarget{definitions}{%
\subsection{Definitions}\label{definitions}}

\emph{Partition} of a set we call a family of disjoint sets that cover
the space.

Reals \(\R\) and positive reals
\(\R_{+} = \{ x \| x \geq 0 \aand x \in \R \}\).

We use symbol \(\leq\) for any partial ordering, \(\union\) for tagged
disjoint union, with \(\mathop{in}_r\) for tagging right and
\(\mathop{in}_l\) for tagging left.

We use \(\union\) as classical set union, and \(\taggedUnion\) for a
union of sets labelled by right label \(\inr{}\) and left label
\(\inl{}\).

\hypertarget{data-space}{%
\subsubsection{Data space}\label{data-space}}

As mentioned before, we will focus on \emph{complete summarizations} of
a given input. By complete we mean that they do not lose information.

We formalize the above notions as \emph{data space}:

\emph{Data space} is a tuple with the following
\((\D, \leq_{\D}, \norm{\_}_{\D}, (\m, e, \diamond))\):

\begin{itemize}
\tightlist
\item
  data carrier set \(\D\);
\item
  information set \(\m\);
\item
  partial ordering representing \emph{information growth}
  \(\leq_{\D}\)\footnote{Of course, \(\leq_{\D}\) is a monoidal
    preorder(Fong and Spivak 2021)};
\item
  measure of data by elements of set \(\m\):
  \(\norm{\_} : \powerset{\D} \rightarrow \m\).
\item
  information fusion monoid
  \(\diamond_{\D} : \m \rightarrow \m \rightarrow \m\), that is additive
  in \(m\);
\item
  neutral element \(e\) of fusion monoid contains is the neutral element
  of \(\m\) that is equal to \(\norm{\emptyset}_{\D}\)
\item
  we expect \emph{information fusion} to be invariant over any partition
  of the carrier set \(\D\);
\item
  \emph{monotonicity} of partial ordering with respect to
  \emph{information fusion}:
\end{itemize}

\begin{enumerate}
\def\labelenumi{(\arabic{enumi})}
\tightlist
\item
  \[ a \leq b \land c \leq d \implies a \diamond c \leq b \diamond d \]
\end{enumerate}

We note that our partial ordering can often be \emph{derived} from well
behaved information monoid by equational laws: \(a \diamond b \leq a\)
and \(a \diamond b \leq b\).

It describes an additive monoid with partial ordering that is consistent
with this monoid. That is because we are interested only in
\emph{complete summarizations} of a data on the given carrier set
\(\D\).

\hypertarget{examples-of-data-spaces}{%
\paragraph{Examples of data spaces}\label{examples-of-data-spaces}}

\begin{itemize}
\tightlist
\item
  \emph{accounting measure} (Ijiri 2018)
\item
  any \emph{mereological space} with a \emph{finitely additive metric},
  which is a family of sets equipped with operations of Boole's algebra,
  and additive metric over these sets(Arntzenius 2004; Barbieri and
  Gerla 2021);
\item
  \emph{typelikes} used for recording union types (Gajda 2022) with a
  partial ordering derived from the information fusion monoid as
  \(a \diamond b \leq a\) and \(a \diamond b \leq b\), and metric
  \(\norm{\_}\) as type inference operation.
\item
  \emph{Banach space} (\emph{metric space}) with set union as
  information fusion.
\end{itemize}

Each set is also equipped with an \emph{identity} data space:

\begin{enumerate}
\def\labelenumi{(\arabic{enumi})}
\setcounter{enumi}{1}
\tightlist
\item
  \[ \begin{array}{rcll}
  \norm{\_}_{\id} & := & \id \\
  \D_{\id} & := & \m_{\id} \\
  \leq & := &  \subset \\
  \diamond_{\id} & := & \union
  \end{array} \]
\end{enumerate}

Another interesting example is \emph{Paccioli algebra} that describes
double-entry accounting:

\begin{enumerate}
\def\labelenumi{(\arabic{enumi})}
\setcounter{enumi}{2}
\tightlist
\item
  \[ \begin{array}{rcl}
  \D_p & := & \R_{+} \times \R_{+} \\
  (d_1, c_1) \diamond (d_1, c_2) & := & (d_1+d_2, c_1+c_2) \\
  \text{partial ordering } (d_1, c_1) \leq_{p} (d_2, c_2)\text{,} & \text{iff} & d_1 \leq d_2$ and $c_1 \leq c_2 \\
  \end{array} \]
\end{enumerate}

This is example, of how we create an additive data space over account
balances, by converting each balance into a pair of strictly positive
credit and debit.

\hypertarget{examples-of-data-spaces-summarizations}{%
\paragraph{Examples of data spaces
(summarizations)}\label{examples-of-data-spaces-summarizations}}

Whenever summarization of data is performed, we should attempt to
perform similar summarization on the fields in the error path, or other
measure of impact counting when these fields are absent.

For each table, we indicate possible summarizations by assigning fitting
monoid to a column, and indicating other columns that may be used for
grouping the results.

For example the table \texttt{order\_details} from typical schema:

\begin{longtable}[]{@{}
  >{\raggedright\arraybackslash}p{(\columnwidth - 8\tabcolsep) * \real{0.3067}}
  >{\raggedright\arraybackslash}p{(\columnwidth - 8\tabcolsep) * \real{0.1200}}
  >{\raggedright\arraybackslash}p{(\columnwidth - 8\tabcolsep) * \real{0.1200}}
  >{\raggedright\arraybackslash}p{(\columnwidth - 8\tabcolsep) * \real{0.1067}}
  >{\raggedright\arraybackslash}p{(\columnwidth - 8\tabcolsep) * \real{0.3467}}@{}}
\caption{Example summarization.}\tabularnewline
\toprule
\begin{minipage}[b]{\linewidth}\raggedright
column
\end{minipage} & \begin{minipage}[b]{\linewidth}\raggedright
type
\end{minipage} & \begin{minipage}[b]{\linewidth}\raggedright
unique
\end{minipage} & \begin{minipage}[b]{\linewidth}\raggedright
result
\end{minipage} & \begin{minipage}[b]{\linewidth}\raggedright
aggregation
\end{minipage} \\
\midrule
\endfirsthead
\toprule
\begin{minipage}[b]{\linewidth}\raggedright
column
\end{minipage} & \begin{minipage}[b]{\linewidth}\raggedright
type
\end{minipage} & \begin{minipage}[b]{\linewidth}\raggedright
unique
\end{minipage} & \begin{minipage}[b]{\linewidth}\raggedright
result
\end{minipage} & \begin{minipage}[b]{\linewidth}\raggedright
aggregation
\end{minipage} \\
\midrule
\endhead
order\_id & integer & unique & set of identifiers & union \\
product\_id & integer & - & set of identifiers & union \\
unit\_price & number & - & number & min, max, avg \\
quantity & number & - & number & sum \emph{(for each unit type)} \\
unit\_price * quantity & number & derived & number & sum \\
\bottomrule
\end{longtable}

Now that we read this example, we also see that reporting monoids
exhibit certain trends:

\begin{itemize}
\tightlist
\item
  OIDs, unique identifiers, dimensions are usually summarized by sets
\item
  number facts are usually summarized by sum, min, max, average
\end{itemize}

It is common to use the same summarization on the same datatype (unit).
Summarizing together different datatypes (units) is usually a mistake,
in which case we use a containing category, for example with count
different types of faulty records instead of summing their values.

Automatic summarization can perform a summarization on all permitted
aggregations in order to detect common patterns.

We can infer, that we may assign a heuristic that for every
summarization of correct data would also summarize incorrect records on
those monoids that can be still used in case of incomplete data.

That should make a reasonable default summarizations for the data
outside the main path (or ``happy path'') of the analysis.

\hypertarget{information-preserving-operations}{%
\subsubsection{Information preserving
operations}\label{information-preserving-operations}}

Information preserving operations can be formulated in a general way as
bidirectional and thus reversible, or as preserving the information
measure.

\textbf{Conjecture:} Reversible operations are always preserving any
specific information measure as defined above.

\hypertarget{partition-of-data-space}{%
\paragraph{Partition of data space}\label{partition-of-data-space}}

Instead of selecting or projecting data from the input, the lossless
operation is to \textbf{partition} the data set into subsets of desired
properties.

For example

For a predicate \(p\), we can mark it as:
\[ (\D_{accepted}, \D_{rejected}) = partition(p, \D_{}) \]

\[ \D_{accepted} = \{ x | x \in \D \aand p(x) \}\]

\[ \D_{rejected} = \{ x | x \in \D \aand p(x) \}\]

\hypertarget{information-preserving-union}{%
\paragraph{Information preserving
union}\label{information-preserving-union}}

Union of two sets is information preserving, if members of each set are
clearly labelled in a way that allows to infer from which input set they
come from. This may be either because they come from \emph{disjoint
sets}, or because they are \emph{explicitly tagged}:
\[ \D_u = \D_1 \taggedUnion \D_2 = \{ \inl{d_1} \| d_1 \in \D_1 \} \union \{ \inr{d_1} \| d_1 \in \D_1 \} \]

In some cases we already have tags that serve the same purpose, that is,
there is an origin function \(f\):

\begin{enumerate}
\def\labelenumi{(\arabic{enumi})}
\setcounter{enumi}{3}
\tightlist
\item
  \[ \exists f. \forall x. f(x) = \begin{cases} \inl{} & x \in \D_1 \\
                                 \inr{} & x \in \D_2 \\
                   \end{cases}
  \]
\end{enumerate}

\hypertarget{information-preserving-function}{%
\paragraph{Information preserving
function}\label{information-preserving-function}}

In case we are interested only in information measures
\(\m_1, .., \m_n\) and \(\m'_1, .., \m'_n\) we can narrow ourselves to
the functions that preserve information for these functions, that is:

Given function \(f : \D \to \D'\) we can compute a corresponding mapping
of \(g : \m_i \to \m'_i\) such that \(g(m(x)) = m'(f(x))\).

\hypertarget{examples-of-information-preserving-functions}{%
\paragraph{Examples of information preserving
functions}\label{examples-of-information-preserving-functions}}

In simplest example, it may occur that \(\D = \D'\) and
\(\m_i = \m'_i\).

Another example would be re-partitioning the dataset with respect to a
predicate, or computing additional quantities for each record.

\textbf{Note:} that aggregation generally only preserve information that
corresponds to their dimension (for example summing up quantities only
preserves this dimension, but not the number of records). Later we will
discuss how to construct aggregations preserving multiple dimensions by
using tensor products of aggregations on individual dimensions.

\hypertarget{derivations-of-complex-data-spaces-from-simpler-ones}{%
\subsubsection{Derivations of complex data spaces from simpler
ones}\label{derivations-of-complex-data-spaces-from-simpler-ones}}

Now we address a derivation of complex data spaces from simpler ones.

\hypertarget{tensor-products}{%
\subsubsection{Tensor products}\label{tensor-products}}

We already saw that \emph{data metric space} can be derived from a
well-behaved \emph{information fusion monoid}.

Now we attempt to assemble \emph{data spaces} from subspaces that
conform to different \emph{data metrics}:

\hypertarget{disjoint-tensor-product}{%
\paragraph{Disjoint tensor product}\label{disjoint-tensor-product}}

\emph{Disjoint tensor product} of two \emph{data spaces}
\(\D_{1} \disjProduct \D_{2}\) is a data space such that:

\begin{itemize}
\tightlist
\item
  data carrier set is \(\D = \D_1 \times \D_2\);
\item
  partial ordering
  \(a \leq c \land b \leq d \implies (a, b) \leq (c, d)\);
\item
  information fusion
  \((a, b) \diamond (c, d) = (a \diamond_{1} c, b \diamond_{2} d)\)
\item
  neutral element \(e = (e_1, e_2)\)
\end{itemize}

\hypertarget{disjoint-tensor-product-1}{%
\subparagraph{Disjoint tensor product}\label{disjoint-tensor-product-1}}

Given two data spaces over the same set, we may want to use composite
\emph{data metric} that represents information conveyed by data metrics
in the both spaces:

\emph{Parallel product} of two \emph{data spaces}
\(\D_{1} \parProduct{} \D_{2}\) is a data space such that:

\begin{itemize}
\tightlist
\item
  carrier set i1s \(\D = \D_1 = \D_2\);
\item
  partial ordering
  \(a \leq c \land b \leq d \implies (a, b) \leq (c, d)\);
\item
  information fusion
  \((a, b) \diamond (c, d) = (a \diamond_{1} c, b \diamond_{2} d)\)
\item
  neutral element \(e = (e_1, e_2)\)
\end{itemize}

\hypertarget{information-reconstruction-and-information-projection}{%
\subparagraph{Information reconstruction and information
projection}\label{information-reconstruction-and-information-projection}}

Given two dataspaces \(\D_1\) and \(\D_2\) such that the data carrier is
the same, but different information carriers \(\m_1\) and \(\m_2\), we
can make a \(\D = \D_1 \reconsProduct \D_2\) by making a tensor product
of information carriers \(\m = \m_1 \parProduct \m_2\) and using induced
partial order and information fusion monoid:

\begin{itemize}
\tightlist
\item
  \(a \leq b\) when both \(\norm{a}_{\D_1} \leq_{\D_1} \norm{b}\) and
  \(\norm{a}_{\D_2} \leq_{\D_2} \norm{b}_{\D_2}\)
\item
  \(m \diamond_{\D_1 \reconsProduct \D_2} n = (m \diamond_{\D_1} n, m \diamond_{\D_2} n)\)
\end{itemize}

For disjoint information carriers, we may call the component dataspaces
\(\D_1\) and \(\D_2\) \emph{projections}:

\[\Proj{\m_1}{\D}=\D_1\]

\[\Proj{\m_2}{\D}=\D_2\]

\hypertarget{products-of-functions}{%
\paragraph{Products of functions}\label{products-of-functions}}

Since we are interested in transformations of \emph{data spaces}, we
denote function products that correspond to the above products on data
spaces:

\[ (f \disjProduct g)(x) \begin{cases} \inr x & \rightarrow f(x) \\
                                  \inl y & \rightarrow g(y) \\
                         \end{cases}
\]

\[ (f \parProduct g)(x) =  (f(x), g(x)) \]

\hypertarget{partial-functions}{%
\paragraph{Partial functions}\label{partial-functions}}

Each partial function \(f : \D \mapsto \m_1\) can be extended to a total
function \(\tot{f} : \D \to \m\) by using the partition of the data
space \(\D\) into two different fragments \(\D_1\) and \(\D_2\):

\[ \tot{f}(a) \begin{cases}
      \inr{f(a)}, & \text{if}\ x \in \D_1 \\
      \inl{a}, & \text{otherwise}
    \end{cases}
\]

That means that \(f : \D \mapsto \m_1 \taggedUnion \compl{D_1}\)

\hypertarget{lookups-and-joins}{%
\paragraph{Lookups and joins}\label{lookups-and-joins}}

Joins may represent lookups of information that may change the impact of
processed records. But it may also be true that even when computing
inner join on correct data, some of the error stream data will have
absent key fields necessary for the join. Because of this, for every
join of correct data we perform a similar \textbf{outer join} on error
path data. Because of this, we need to \textbf{always} perform outer
join on error path data, with both records that are on the correct path
and error records

We limit information preserving operations to outer join with a tagging
of results to left, right, and inner output. Inner output consists of
tuples matching by the join condition \(c\), left part consists of
tuples from \(\D_1\) that were not matched by any of \(\D_2\), and right
part consists of tuples from \(\D_2\) that were not matched by any of
\(\D_1\).

\[\D_1 \join_{c} \D_2 = (\D_{\textit{left}}, \D_{\textit{inner}}, \D_{\textit{right}})\]

For a predicate \(p\), we can mark it as:
\[ \D_{inner}) = \{ (x, y) | x \in \D_1 \aand y \in \D_2 \aand p(x, y) \} \]

\[ \D_{left} = \{ x | x \in \D_1 \aand \not \exists_{y \in \D_2} p(x, y) \}\]

\[ \D_{right} = \{ x | x \in \D_1 \aand \not \exists_{y \in \D_2} p(x, y) \}\]

Note that join condition \(c\) is a filter on a cartesian product of
\(\D_1\), and \(\D_2\) that only depends on the join columns (usually
keys).

\hypertarget{value-lookup-example}{%
\subparagraph{Value lookup example}\label{value-lookup-example}}

We take an example of looking up current price of the commodity in the
table including \texttt{product}, and \texttt{quantity} of the position.
The reference table is keyed by \texttt{product} and has
\texttt{description}, and \texttt{current\_price}. Normal lookup is
inner join when we have guaranteed match of \texttt{product\_id}, which
we convert to outer join, where left side corresponds to
\texttt{product\_id} entries that are \texttt{missing} from the
reference table, and right side corresponds to \texttt{product\_id}
entries in reference table that are \texttt{unused}.

Either left and right side are summarized into error feedback for the
left table, and \texttt{unused} feedback for the right table to assure
that records are never lost.

This allows us to easily find outdated references, or misspellings like
\texttt{oli} instead of \texttt{oil} and report them correctly as a
summary row with a count of 14 rows, and 5 units with unknown product
name of \texttt{Nut\ oli} that was a result of mistyping.

\hypertarget{value-lookup-example-1}{%
\subparagraph{Value lookup example}\label{value-lookup-example-1}}

We take an example of looking up current price of the commodity in the
table including \texttt{product}, and \texttt{quantity} of the position.
The reference table is keyed by \texttt{product} and has
\texttt{description}, and \texttt{current\_price}. Normal lookup is
inner join when we have guaranteed match of \texttt{product\_id}, which
we convert to outer join, where left side corresponds to
\texttt{product\_id} entries that are \texttt{missing} from the
reference table, and right side corresponds to \texttt{product\_id}
entries in reference table that are \texttt{unused}.

Either left and right side are summarized into error feedback for the
left table, and \texttt{unused} feedback for the right table to assure
that records are never lost.

This allows us to easily find outdated references, or misspellings like
\texttt{oli} instead of \texttt{oil} and report them correctly as a
summary row with a count of 14 rows, and 5 units with unknown product
name of \texttt{oli}.

\hypertarget{columnar-projection-and-row-format}{%
\paragraph{Columnar projection and row
format}\label{columnar-projection-and-row-format}}

Since we enforce principle of never discarding the data that can be
relevant to debugging effort, we cannot simply discard fields with
\emph{projection} operator. Because of this, our records contain two
types of fields:

\begin{itemize}
\tightlist
\item
  fields \textbf{relevant} to the final output
\item
  fields \textbf{irrelevant} to final output and only kept for the
  debugging purposes
\end{itemize}

For each record, we have a single value for each \textbf{relevant}
field, but possibly a set of values for each \textbf{irrelevant} field
to take account of discarding duplicate \textbf{relevant} records. That
means that we have a subrelation for each record that keeps only
irrelevant fields.

This allows us to implement a \emph{lossless projection} operation that
moves relevant fields into irrelevant field set.

Note that the projection operation is executed the same way on both
\emph{correct relation} and \emph{error tracking} relation that pairs
with it.

\hypertarget{per-row-data-enrichment-operations}{%
\subsection{Per-row data enrichment
operations}\label{per-row-data-enrichment-operations}}

In SQL syntax, it is common to add a computed field to a query from a
single table:

\begin{Shaded}
\begin{Highlighting}[]
\KeywordTok{select} \OtherTok{"Purchase price"}\OperatorTok{*}\OtherTok{"Amount"} \KeywordTok{as} \OtherTok{"Total price"}
  \KeywordTok{from}\NormalTok{ items;}
\end{Highlighting}
\end{Shaded}

We call these operations \emph{data enrichment} and can execute them on
all rows. However, in order to maximize debugging potential, we propose
a methodology of automatically shifting computing additional fields as
early in the pipeline of operations as possible. That allows us to
provide more information. After the record leaves the \emph{correct
stream} and goes into \emph{error stream}, we no longer enrich it with
new field values relevant to the main computation in order to prevent
\emph{cascading faults}.

We can only enrich these records with more \emph{error-specific}
information, like adding a description of error found, and estimating
the impact of this information.

That is why beside \texttt{fmap} operation on \emph{correct data
stream}, we also allow \texttt{emap} operation on the error stream that
enriches error information. We can optionally schedule this information
to be computed in advance when the stage of pipeline is reached, and
before any records are added to the error stream:

\begin{verbatim}
pipeline = do
  forM filename -> do
    emap (@"Filename" :-> filename) $ do
      ...
\end{verbatim}

\hypertarget{putting-algebra-together}{%
\paragraph{Putting algebra together}\label{putting-algebra-together}}

The final step is to use these operation in a \emph{relevant} way: that
means that all data spaces need to be either included in the final
aggregation, or used by data spaces that contribute to it. Just like in
linear logic we are not allowed to use certain variables more than once,
in the data space algebra we are not allowed to \emph{forget} a
variable.

\hypertarget{translating-relational-algebra}{%
\subsection{Translating relational
algebra}\label{translating-relational-algebra}}

Relational algebra is a standard algebra used as basis for describing
database operations, and analytics pipelines.

These primitives are usually implemented on sets of rows (records of the
same shape) called \textbf{relations}. However, they can equally well be
implemented on multisets or lists of records\footnote{In which case,
  usually there are an additional operation for sorting. and extracting
  subset of unique, distinct records.}

It uses the basic primitives of:

\begin{itemize}
\tightlist
\item
  projection which discards a set of columns, and keeps only the columns
  relevant for the further processing
\item
  selection which selects a subset of records
\item
  rename which renames some fields to others
\item
  cartesian product (cross product) of two relations, with outputs being
  a list of all possible pairs of rows from both inputs
\item
  set operations of union, difference, and intersection
\end{itemize}

Note that all joins, intersection, and difference operators that are
technically implemented as outer joins with a post-processing step.

For convenience we also provide lambda or map operation that computes a
new field (or fields) from the others (logically redundant, but provided
as convenience).

\begin{longtable}[]{@{}
  >{\raggedright\arraybackslash}p{(\columnwidth - 4\tabcolsep) * \real{0.1258}}
  >{\raggedright\arraybackslash}p{(\columnwidth - 4\tabcolsep) * \real{0.2264}}
  >{\raggedright\arraybackslash}p{(\columnwidth - 4\tabcolsep) * \real{0.6478}}@{}}
\caption{Translation of operations from relational algebra and SQL to
data space algebra.}\tabularnewline
\toprule
\begin{minipage}[b]{\linewidth}\raggedright
Relational algebra
\end{minipage} & \begin{minipage}[b]{\linewidth}\raggedright
SQL
\end{minipage} & \begin{minipage}[b]{\linewidth}\raggedright
Data space algebra
\end{minipage} \\
\midrule
\endfirsthead
\toprule
\begin{minipage}[b]{\linewidth}\raggedright
Relational algebra
\end{minipage} & \begin{minipage}[b]{\linewidth}\raggedright
SQL
\end{minipage} & \begin{minipage}[b]{\linewidth}\raggedright
Data space algebra
\end{minipage} \\
\midrule
\endhead
Projection & \texttt{SELECT\ field\ FROM\ ...} & extraction \\
selection & \texttt{WHERE\ ...} or \texttt{HAVING\ ...} & partition \\
rename & \texttt{WHERE\ ...} or \texttt{HAVING\ ...} & mapping
function \\
cross product & \texttt{SELECT\ ...\ FROM\ table1,\ table2} & disjoint
tensor product \\
outer join & \texttt{OUTER\ JOIN} & partition induced by NULL on each
table, then parallel tensor product \\
natural join & \texttt{INNER\ JOIN} & parallel tensor product on
partition induced by non-NULL for each table \\
union & \texttt{UNION} & disjoint tensor product (tagged union), then
partition removing duplicates, and mapping removing tags \\
multiset union & \texttt{UNION\ ALL} & tagged union, then
information-preserving function that removes tags \\
difference & \texttt{MINUS} & partition on the first table that puts
shared elements in the error set \\
intersection & \texttt{INTERSECT} & partition on each table that puts
shared element in the output set \\
\bottomrule
\end{longtable}

Using these operations, we can also implement joins Additionally there
are joins, and aggregation operations provided.

Joins are operations that take two relations and produce rows that are
contains bigger rows, that are merged from both inputs.

Additionally we also add the aggregate operations:

\begin{itemize}
\tightlist
\item
  groupings which produce a relation with records that are subrelations
  with fields from the same subgroup
\item
  aggregation operations which take any monoid operation over a given
  field, and produce rows that are monoid sum over all records in a
  relation
\end{itemize}

The set of operations above are widely acknowledged as sufficient to
provide most of ETL and data aggregation needs. Whenever a new database
or data analytics operation is proposed, it is customary to compare it
to basic relational algebra operations. Thus we will propose a set of
operations that allow us to get the same results as relational algebra
expressions.

\hypertarget{discussion}{%
\section{Discussion}\label{discussion}}

\hypertarget{error-aggregates-vs-error-estimates}{%
\subsection{Error aggregates vs error
estimates}\label{error-aggregates-vs-error-estimates}}

Use of error-tracing relational algebra allows us to easily make general
assertions on how the individual records impact processing (since each
total summarization must be sensitive to all records). When we expect a
given record to be 100\% correct, we just assert its fields are included
in a correct recordset at the end of correct path.

If we expect a given record to be discarded, we may just assert that it
is present in the error set.

That also means that we may use a failing assertion to quickly point to
all possible code sites or data inputs that may need to be changed in
order to fix the issue.

\hypertarget{replacing-relational-algebra-in-analytics}{%
\subsection{Replacing relational algebra in
analytics}\label{replacing-relational-algebra-in-analytics}}

We proposed a replacement of relational algebra with set of operations
that allow for the same computation, but structured in a way that assure
all data is summarized and included. We justify it by analogy to
accounting theory, and thus coin the name ``data accounting'' for the
process of complete summarization of a data set in order to assure
conformance to statististical and analytical best practice. We described
the process of translation of classical relational algebra into this new
``data space algebra''. This algebra allows for reversible
transformation of data, and preservation of mathematical principles of
accounting while aggregating the data. Implementation of ``data space
algebra'' operations on top of traditional SQL is shortly described.

\hypertarget{conclusion}{%
\section{Conclusion}\label{conclusion}}

We also suggest a common summarization mechanism for all well-defined
columns within the error table in order to facilitate finding of
patterns in the errors.

We provide data space algebra as an alternative to conventional
relational algebra for the purpose of more robust reporting, and suggest
using accounting principles for more disciplined data analysis. We
shortly described how to convert traditional relational algebra
expressions so that there one of the outputs gives the same information
as relational query. Naturally for full accounting of input data we also
require additional information to be summarized in order to assure that
all data is accounted for.

\hypertarget{references}{%
\section*{Bibliography}\label{references}}
\addcontentsline{toc}{section}{Bibliography}

\hypertarget{refs}{}
\begin{CSLReferences}{1}{0}
\leavevmode\vadjust pre{\hypertarget{ref-differentiable-computing}{}}%
Abadi, Martin, and Gordon D. Plotkin. 2019. {``A Simple Differentiable
Programming Language.''} \emph{Proc. ACM Program. Lang.} 4 (POPL).
\url{https://doi.org/10.1145/3371106}.

\leavevmode\vadjust pre{\hypertarget{ref-mereology-measure}{}}%
Arntzenius, Frank. 2004. {``Gunk, Topology and Measure.''} In
\emph{Oxford Studies in Metaphysics: Volume 4}, edited by Dean
Zimmerman. Oxford University Press.

\leavevmode\vadjust pre{\hypertarget{ref-axiomatic-accounting-mattesich}{}}%
Balzer, Wolfgang, and Richard Mattessich. 1991. {``An Axiomatic Basis of
Accounting: A Structuralist Reconstruction.''} \emph{Theory and
Decision} 30 (3): 213--43.

\leavevmode\vadjust pre{\hypertarget{ref-bidirectional-computing}{}}%
Bancilhon, François, and Nicolas Spyratos. 1985. {``Algebraic Versus
Probabilistic Independence in Data Bases.''} In \emph{Proceedings of the
Fourth {ACM} {SIGACT-SIGMOD} Symposium on Principles of Database
Systems, March 25-27, 1985, Portland, Oregon, {USA}}, 149--53. {ACM}.
\url{https://doi.org/10.1145/325405.325424}.

\leavevmode\vadjust pre{\hypertarget{ref-mereological-space}{}}%
Barbieri, Giuseppina, and Giangiacomo Gerla. 2021. {``Defining Measures
in a Mereological Space (an Exploratory Paper).''} \emph{Logic and
Logical Philosophy} 31 (1): 57--74.
\url{https://doi.org/10.12775/LLP.2021.005}.

\leavevmode\vadjust pre{\hypertarget{ref-bcbs-239}{}}%
Basel Committee on Banking Supervision. 2013. {``BCBS 239: Principles
for Effective Risk Aggregation and Risk Reporting.''}
\url{https://www.bis.org/publ/bcbs239.htm}.

\leavevmode\vadjust pre{\hypertarget{ref-reversible-computing}{}}%
Bennett, C. H. 1973. {``Logical Reversibility of Computation.''}
\emph{IBM Journal of Research and Development} 17 (6): 525--32.
\url{https://doi.org/10.1147/rd.176.0525}.

\leavevmode\vadjust pre{\hypertarget{ref-view-update}{}}%
Chen, Haitao, and Husheng Liao. 2011. {``A Survey to View Update
Problem.''} \emph{International Journal of Computer Theory and
Engineering} 3: 23--31.

\leavevmode\vadjust pre{\hypertarget{ref-paccioli-algebras}{}}%
Ellerman, David. 2007. {``Double-Entry Accounting: The Mathematical
Formulation and Generalization.''} \emph{SSRN Electronic Journal},
December. \url{https://doi.org/10.2139/ssrn.1340619}.

\leavevmode\vadjust pre{\hypertarget{ref-monoidal-preorder}{}}%
Fong, Brendan, and David I. Spivak. 2021. {``Symmetric {Monoidal}
{Preorders}.''} Massachusetts Institute of Technology.

\leavevmode\vadjust pre{\hypertarget{ref-data-science-conference}{}}%
Gajda, Michal Jan. 2020a. {``Agile Data Pipelines: ETL for 2020.''} Data
Science Europe. 2020. \url{https://youtu.be/aHAc8ght9Gw}.

\leavevmode\vadjust pre{\hypertarget{ref-skills-matters}{}}%
---------. 2020b. {``Agile Generation of Cloud API Bindings with
Haskell.''} Haskell.Love. 2020.
\url{https://skillsmatter.com/skillscasts/14905-agile-functional-data-pipeline-in-haskell-a-case-study-of-multicloud-api-binding}.

\leavevmode\vadjust pre{\hypertarget{ref-towards-perfect-union}{}}%
---------. 2022. {``{Towards a more perfect union type}.''}

\leavevmode\vadjust pre{\hypertarget{ref-ijiri-book}{}}%
Ijiri, Yuji. 1975. \emph{Theory of Accounting Measurement}. 10. American
Accounting Association.

\leavevmode\vadjust pre{\hypertarget{ref-ijiri}{}}%
---------. 2018. {``Axioms and Structures of Conventional Accounting
Measurement.''} \emph{Accounting, Economics, and Law: A Convivium} 8
(1): 20170057.
https://doi.org/\url{https://doi.org/10.1515/ael-2017-0057}.

\leavevmode\vadjust pre{\hypertarget{ref-distributed-error-tracking}{}}%
Julian, Mike. 2017. {``Practical Monitoring.''}

\leavevmode\vadjust pre{\hypertarget{ref-levitt-md}{}}%
Levitt, Michael. 1983. {``Molecular Dynamics of Native Protein: I.
Computer Simulation of Trajectories.''} \emph{Journal of Molecular
Biology} 168 (3): 595--617.
https://doi.org/\url{https://doi.org/10.1016/S0022-2836(83)80304-0}.

\leavevmode\vadjust pre{\hypertarget{ref-monotonic-database-computation}{}}%
Ross, Kenneth A, and Yehoshua Sagiv. 1997. {``Monotonic Aggregation in
Deductive Databases.''} \emph{Journal of Computer and System Sciences}
54 (1): 79--97.
https://doi.org/\url{https://doi.org/10.1006/jcss.1997.1453}.

\leavevmode\vadjust pre{\hypertarget{ref-maths-of-accounting-as-question}{}}%
Warsono, Sony, Muhammad Ridha, and Arif Darmawan. 2009. {``Mathematics
in Accounting as a Big Unanswered Question.''} \emph{SSRN Electronic
Journal}, July. \url{https://doi.org/10.2139/ssrn.1439084}.

\leavevmode\vadjust pre{\hypertarget{ref-railroad-programming}{}}%
Wlaschin, Scott. 2012. {``F\# for Fun and Profit.''} 2012.
\url{https://swlaschin.gitbooks.io/fsharpforfunandprofit/content/posts/recipe-part2.html}.

\end{CSLReferences}

\bibliographystyle{eptcs}

\bibliography{data-accounting.bib}

\end{document}